\documentclass{article}[11pt]

\usepackage{authblk}

\usepackage[square,numbers]{natbib}
\usepackage[a4paper]{geometry}
\usepackage{microtype}
\usepackage{siunitx}
\usepackage{amsmath}
\usepackage{amssymb}
\usepackage{graphicx}

\usepackage{setspace}
\providecommand{\keywords}[1]
{
  \small	
  \textbf{\textit{Keywords---}} #1
}

\title{The role of grain-environment heterogeneity in normal grain growth: a stochastic approach}
\author[1,*]{Thomas Breithaupt}
\author[1]{Lars N. Hansen}
\author[2,3]{Srikanth Toppaladoddi}
\author[1]{Richard F. Katz}
\affil[1]{Department of Earth Sciences, University of Oxford, South Parks Road, Oxford, OX1 3AN, UK}
\affil[2]{Department of Physics, University of Oxford, Oxford OX1 3PU}
\affil[3]{Mathematical Institute, University of Oxford, Oxford OX2 6GG}
\affil[*]{Corresponding author: thomas.breithaupt@univ.ox.ac.uk}

\date{}

\begin{document}
\maketitle

\section*{Abstract}

The size distribution of grains is a fundamental characteristic of polycrystalline solids. In the absence of deformation, the grain-size distribution is controlled by normal grain growth. The canonical model of normal grain growth, developed by Hillert, predicts a grain-size distribution that bears a systematic discrepancy with observed distributions. To address this, we propose a change to the Hillert model that accounts for the influence of heterogeneity in the local environment of grains. In our model, each grain evolves in response to its own local environment of neighbouring grains, rather than to the global population of grains. The local environment of each grain evolves according to an Ornstein-Uhlenbeck stochastic process. Our results are consistent with accepted grain-growth kinetics. Crucially, our model indicates that the size of relatively large grains evolves as a random walk due to the inherent variability in their local environments. This leads to a broader grain-size distribution than the Hillert model and indicates that heterogeneity has a critical influence on the evolution of microstructure.

\keywords{Grain growth; Grain size distribution; Monte Carlo simulation; Heterogeneity}

\pagebreak

\section{Introduction}

Many properties of engineered and natural materials are controlled by the size distribution of their constituent grains. Accurately modelling the evolution of the distribution of grain size is critical to understanding these properties, and is a prerequisite to modelling processes such as strain localisation due to grain-size reduction \cite{rozel2011thermodynamically}. The fundamental process acting on grain size is normal grain growth; to confidently predict the evolution of a population of grains undergoing deformation, we must first predict the grain-size distributions that result from normal grain growth. Normal grain growth is driven by the tendency to minimise the total surface energy of the grains. In observations of normal grain growth from laboratory experiments and numerical simulations, the mean grain size grows as the square root of time, and the distribution of grain size normalized by the mean is constant \cite{rios2018}. The latter property is termed statistical self-similarity. The classical, mean-field theory of grain growth due to Hillert \cite{hillert1965} assumes that grains are spherical and grow at a rate that depends on each grain's self-curvature and a global, mean-field curvature. Hillert's model achieves the observed square-root of time kinetics but predicts a self-similar distribution that is inconsistent with observations \cite{rios2018, rios2006comparison, rios2006self}. Despite this inconsistency, analysis of the thermodynamics \cite{fischer2003thermodynamic, kertsch2016modelling} and grain topology \cite{rios2006topological} have provided support for Hillert kinetics, and the Hillert model forms the basis for many models of normal grain growth  \cite[e.g.,][]{brown1989new, pande1990n2, marthinsen1996influence, zollner2006three}. We take the kinetics of the Hillert model as our starting point.

We modify the Hillert model by replacing the mean-field curvature with a local curvature, defined as the mean curvature over the local environment of each grain. The local environment of a grain is the set of its nearest-neighbour grains, i.e., the grains that it exchanges mass with. The local environment of a grain interacts with a much larger set of grains, many of which are screened from the grain by its neighbours.  From the perspective of individual grains, the local curvature will undergo a random walk, providing a source of noise to the grain-size kinetics. To capture this effect, we introduce a stochastic term into our model with an amplitude determined by the global statistics of local environments. Our approach differs from other stochastic models because the amplitude of our noise is determined through the hypothesis of a specific physical source \cite{pande1990n2, pande2010self, pande2020beyond}.

The local curvature is defined as an average grain curvature over the local environment, which is itself a small sample of the global population of grains. Hence, the distribution of local curvatures will approximate a normal distribution. We therefore model local curvatures using an Ornstein-Uhlenbeck process, because realisations of this process are normally distributed \cite{uhlenbeck1930theory, wang1945theory}. We solve the coupled stochastic differential equations for grain radius and local environment for a large ($N \approx 10^6$) set of grains over a sufficiently long period of time for the set to reach statistical self-similarity. We obtain grain-size distributions that are in agreement with observations, indicating that heterogeneity in the local environments of grains plays a key role in normal grain growth.

\subsection{Classical models of grain growth}

In his canonical model, Hillert \cite{hillert1965} assumed the simplest possible expression for the grain-growth rate that gives the required behaviour: that small grains shrink and large grains grow. As a result, Hillert's theory does not explicitly incorporate key physical principles such as the geometric requirement of equilibrium dihedral angles and the topological requirement of filling space.
Nevertheless, Hillert's theory forms the basis of many models of normal grain growth \cite[e.g.][]{brown1989new, pande1990n2, marthinsen1996influence, zollner2006three}. More recently, Hillert's model has gained support from topological and thermodynamic perspectives. Rios \& Glicksman \cite{rios2006self} develop a special set of polyhedra that possess equilibrium dihedral angles and are space filling. By assuming that every grain can be approximated by a member of this set, they develop equations for the rate of change of grain size and the number of grain faces. Ultimately they demonstrate that, given certain assumptions about grain topology, their model is equivalent to Hillert's, validating Hillert's approach. Similarly, distinct thermodynamic approaches derive a variation of the Hillert model \cite{fischer2003thermodynamic, kertsch2016modelling} that reduces to the classical Hillert model under the assumption that the mobilities of all grain boundaries are given by a single constant. However, broadly recognised discrepancies between observations and Hillert's distribution indicate that the theory does not provide an adequate approximation to the grain-growth rate \cite{rios2018, rios2006comparison, rios2006self}.

Modifications to Hillert's model centre on the introduction of adjustable parameters or stochastic terms. Brown \cite{brown1989new} suggested relaxing a stability argument that Hillert had applied, and thereby introduced an adjustable parameter into the model. Rios \cite{rios1999comparisonBROWN} applied this approach to derive a one-parameter family of distributions, some of which are in agreement with observed distributions \cite{rios2006comparison}. However, the stability of these distributions and the physical justification for Brown's approach are a matter of debate \cite{rios1999comparisonBROWN}, and the selection of the best-fit Rios distribution is essentially a ``phenomenological procedure'' \cite{rios2001comparison}. In contrast, Marthinsen et al. \cite{marthinsen1996influence} used numerical simulations to guide modification of Hillert's theory. In their simulations, an inverse correlation develops between a grain's size and its local environment. This correlation develops because large grains grow by taking mass away from their neighbours, shrinking them. Marthinsen et al. \cite{marthinsen1996influence} regress this correlation and adapt Hillert's model into a correlation-field theory of normal grain growth. However, their predicted distribution is too negatively skewed to match observations. Streitenberger \cite{streitenberger1998generalized} furthered this approach by deriving a constraint on the linear fit between the local environment and grain size. Using only this constraint and the intercept of the relationship in \cite{marthinsen1996influence}, Streitenberger predicts a grain-size distribution that is in agreement with the distribution observed by Marthinsen et al. However, the gradient Streitenberger infers is much higher than the gradient of the correlation observed by Marthinsen et al., which Streitenberger attributes to topological constraints not accounted for in the model.

Alternatively, Hillert's model can be augmented with stochastic terms. By considering the statistical divergence between two solutions for the grain-size distribution, Pande \& Rajagopal \cite{pande2001uniqueness} argued that under deterministic grain-size kinetics, arbitrary initial distributions cannot converge to a unique self-similar distribution. Consequently, grain-size kinetics must incorporate a stochastic term to reproduce this key feature of normal grain growth. Typically, an additive white-noise term is appended to the Hillert kinetics \cite{pande1990n2, pande2008grain}. The strength of the noise introduced by this term is left as an adjustable parameter, and so its presence is not directly attributed to any one process. However, it has been suggested that the effect of heterogeneity in the local environments of grains provides a source of noise \cite{pande1991n3, pande2020beyond}. In this case, the self-similar grain-size distribution can be found by solving the Fokker-Planck equation associated with the proposed grain-size kinetics. Pande \& Moser \cite{pande2020beyond} seek an approximate analytical solution and find that it is equivalent to a previously derived distribution from a modified mean-field theory \cite{streitenberger2006effective} and that is in agreement with observed distributions.

The application of white noise to normal grain growth is not without controversy, however. Implicit in the assumption of white noise is that the process causing fluctuations in grain size occurs on a much shorter timescale than the process driving deterministic evolution of grain size \cite{mullins1998grain}. Since neighbouring grains should change their size on the timescale of grain-size evolution itself, Mullins argued that fluctuations in grain-growth rate within these models cannot arise from the neighbouring grains \cite{mullins1998grain}. In contrast, Pande \& Moser \cite{pande2020beyond} have argued that any fluctuation-causing process that affects a single grain will ultimately affect the growth rate of every grain in the system. As the number of grains becomes large, the timescale of fluctuations becomes arbitrarily small. White noise can therefore be applied to model normal grain growth under the assumption that the spatial influence of each fluctuation is arbitrarily large.

\section{Our model}

We propose a modification to Hillert's theory that is associated with the mean-field curvature. We adjust the mean-field curvature by a local term that is specific to each grain. This local term accounts for the difference between the mean-field environment and the local environment of each grain. Our approach differs from previous modifications because the local environment of each grain is allowed to evolve independently. We hypothesise that the differential of the size of the  $i$\textsuperscript{th} grain $\text{d}{R_i}$ is given by 
\begin{align}
    \text{d}{R_i}(t) = \alpha \left( \frac{1}{R_m(t) + S_i(t)} - \frac{1}{{R_i}(t)} \right) \text{d}t,
\label{eqn:our_model}
\end{align}
where $\alpha$, which has the units of diffusivity, is the product of the surface energy per unit area, the grain-boundary mobility, and a geometric factor of order unity. We assume that the surface energies and the mobilities of all grain boundaries are identical, an assumption that has been previously called the uniform-boundary model \cite{rios2004irreversible}. Consequently, $\alpha$ is a constant for all grains. $R_m(t)$ is the mean-field radius, ${S_i}$ is the deviation of the local environment from the mean field, and $t$ is time.  Variables labelled with a subscript $i$ are local quantities, specific to the {\textit{i}}\textsuperscript{th} individual grain. We develop a model for the evolution of $S_i$ below. Hillert's theory can be recovered from Eq. \eqref{eqn:our_model} by setting ${S}_i(t)$ equal to zero. We suggest that Eq. \eqref{eqn:our_model} better reflects the physics of grain growth at the grain scale, for which the size of a grain relative to its neighbours plays a role in determining grain-growth rates.

The mean-field radius is calculated by requiring the total mass of the set of grains to be conserved. Since density is constant for a single phase, this is equivalent to conservation of total grain volume. Assuming that each grain is spherical, the differential of grain volume $\text{d}{V_i}$ is given by
\begin{align}
    \text{d}{V_i}(t) = 4 \pi \alpha  \left( \frac{{R_i}^2}{R_m(t) + {S_i}(t)} - {R_i}\right) \text{d}t.
    \label{eqn:vol_increments}
\end{align}
Total volume conservation is enforced by requiring that the sum of all volume increments given by Eq. \eqref{eqn:vol_increments} vanishes, or equivalently, that the expectation value of $\text{d}{V_i}$ vanishes. For the expectation of $\text{d}{V_i}$ to vanish, it must be true that
\begin{align}
    \text{E}\left[ \frac{{R_i}(t)^2}{R_m(t) + {S_i}(t)} \right] - \text{E}[{R_i}(t)] = 0,
    \label{eqn:our_volume}
\end{align}
where $E[\cdot]$ is the expectation defined over the complete set of grains. Eq. \eqref{eqn:our_volume} implicitly determines the mean-field radius $R_m$. In Hillert's theory, ${S_i}$ is zero and the mean-field radius is $\text{E}[{R_i}^2]/\text{E}[{R_i}]$.

The local environment radius $R_m + S_i$ represents the interactions of a grain with its neighbours. We assume that these interactions are exactly governed by the Hillert model and are volume conserving. However, since a grain's interactions are confined to its neighbourhood only, the average over all grains used to compute the mean-field radius in the Hillert model ($\text{E}[R_i^2]/\text{E}[R_i]$) is, in our model, restricted to the neighbourhood of the grain only. Consequently, the local environment radius is here given by ${\text{E}_i}[{R_i}^2]/{\text{E}_i}[{R_i}]$, where ${\text{E}_i}[\cdot]$ is the local mean---the expectation defined over the set of a grain and its neighbours---and is local to an individual grain. The distribution of local environment radii is a sampling distribution of this statistic from the global distribution of grain size. We make the simplifying assumption that this sampling distribution is normally distributed and has a variance given by the small-sample distribution of the mean of ${R}_i$, under the central limit theorem. Therefore, the deviation ${S_i}$ from the mean-field radius is normally distributed as
\begin{align}
    {S}_i \sim \mathcal{N}\left(0, \frac{1}{n} \text{Var}[{R_i}] \right),
    \label{eqn:st_statistics}
\end{align}
with a variance $\text{Var}[{R_i}]/n $, where $n$ is the number of grains in the small-sample set of a grain and its neighbours. The number of grains surrounding a given grain will depend on the size of that grain compared to the size of those surrounding it; grains that are small should have comparatively few neighbours. Introducing this dependence would couple the evolution of a grain's local environment ${S_i}$ to its size ${R_i}$ in a manner that goes beyond the sensitivity to the whole set of grains implied by Eq. \eqref{eqn:st_statistics}. To avoid this complication, we make the simplifying assumption that $n$ is a constant for all grains.

The grains comprising the local environment of an individual grain interact more broadly than does the individual grain itself. In particular, they exchange mass with a much larger set of grains. Heterogeneity in the size of grains within this larger set will drive heterogeneity in exchanges of mass. Consequently, local environments that are initially similar may evolve along very different (and random) trajectories. We make the simplifying assumption that these random walks can be modelled as being Brownian. Consequently, the evolution of $S_i$ is modelled by a stochastic differential equation with a random term $\Omega \, \text{d}W(t)$, for which $\Omega^2/2$ is a diffusion coefficient and $W(t)$ is the standard Wiener process. $\Omega$ quantifies the heterogeneity in the evolution of local environments, which is a consequence of heterogeneity in the broader set of grains that exchange mass with the local environment.

Furthermore, as the grains that comprise the local environment evolve, some may shrink to zero size. Once this occurs, they may be replaced by grains from beyond the local environment. The introduction of replacement grains from the global set of grains will tend to pull the local environment radius $R_m + S_i$ towards the mean-field radius $R_m$. A detailed description of this process would involve modelling the episodic loss of grains from the local environment and the grains that replace them. Instead, we make the simplifying assumption that the replacement of grains can be approximated by the continuous decay of the deviation $S_i$ between the local-environment radius and the mean-field radius on the timescale of removal, which is the timescale of normal grain growth $\tau$. Mathematically, this can be represented by $-(S_i/\tau) \text{d}t$, where

\begin{equation}
    \tau(t) = 2 \frac{\text{E}[{R}_i]^2}{\alpha}.
\end{equation}

As the timescale $\tau$ increases, the decay of $S_i$ will be retarded. Physically, this reduced decay rate corresponds to a reduction in the rate of introduction of replacement grains into any given local environment that occurs as grain growth and the associated removal of grains slows.
 
Combining both of these components, we obtain an Ornstein-Uhlenbeck process

\begin{equation}
\text{d}{S}_i(t) = - \frac{{S_i}(t)}{\tau(t)} \text{d}t + \Omega(t) \text{d}W(t).
\label{eqn:SDE_St}
\end{equation}

The ensemble statistics of $S_i$ generated by Eq. \eqref{eqn:SDE_St} are normal and, provided that the initial mean of $S_i$ is zero, consistent with the statistics given by Eq. \eqref{eqn:st_statistics}. $\Omega$ can be determined by requiring that $S_i$ has the variance given by Eq. \eqref{eqn:st_statistics}. Consequently, $\Omega$ will depend on the variance of the grain-size distribution, which is a measure of heterogeneity in grain size. Since the variance of the grain-size distribution is a function of time, we let $\Omega$ vary with time.

Normal grain growth is a spontaneous irreversible process \cite{rios2004irreversible}. For the proposed model to be thermodynamically admissible, by the second law of thermodynamics, the rate of entropy production must be positive. In our notation, this is equivalent to the statement that $\text{E}[R_i \dot{R}_i]$ is negative \cite{fischer2003thermodynamic}. This condition is satisfied by the Hillert model because the mean-field radius $R_m$ exceeds the mean radius $\text{E}[R_i]$. We have assumed that local interactions within a grain neighbourhood are governed by the Hillert model and therefore in our model, the entropy production rate is positive within every grain neighbourhood. Moreover, we have approximated the distribution of local environment radii as a normal distribution (Eq. \eqref{eqn:st_statistics}). Under this approximation, positivity of entropy production holds provided that the standard deviation of $S_i$ is sufficiently small relative to $R_m$. This condition is satisfied for all of the results we have obtained.

To remove material-specific properties from the model, we non-dimensionalise variables using the initial mean grain size $\text{E}[{R_i}(0)]$ as the scale for grain size and $2 \text{E}[{R}_i(0)]^2/\alpha$ as the scale for time. This timescale is chosen such that under the Hillert model, once the self-similar state is achieved, the square of the mean-field radius is a linear function of time with unit gradient. Equations \eqref{eqn:our_model}, \eqref{eqn:our_volume}, and \eqref{eqn:SDE_St} take the non-dimensional form
\begin{align}
    \text{d}{R}_i' &= 2 \left( \frac{1}{R_m'(t') + {S}_i'} - \frac{1}{{R_i}'} \right) \text{d}t', \label{eqn:nondimR} \\
    \text{d}{S_i}' &= - \frac{{S_i}'}{\text{E}[{R_i}']^2} \text{d}t' + \Omega'(t') \text{d}W(t'), \label{eqn:nondimS} \\
    0 &= \text{E}\Big[ \frac{{R_i}'^2}{R_m'(t') + {S_i}'} \Big] - \text{E}[{R_i}'], \label{eqn:nondimRm}
\end{align}
where a prime denotes a non-dimensional variable. From this point, we drop primes and maintain pre-scaled notation for the rest of the paper. Equations \eqref{eqn:nondimR} and \eqref{eqn:nondimS} allow calculation of the time evolution of ${R_i}$ and ${S_i}$ provided that $R_m(t),\;\text{E}[{R_i}(t)]$, and $\Omega(t)$ are known. 

To find the self-similar distribution of grain size that develops, we take a Monte Carlo approach of simulating a large set of $({R}_i, {S}_i)$ pairs as a function of time. For each pair, we integrate equations \eqref{eqn:nondimR} \& \eqref{eqn:nondimS} forward in time with the Euler-Maruyama scheme, using code developed in the framework of the Portable, Extensible Toolkit for Scientific Computation (PETSc) \cite{petsc-user-ref, petsc-efficient}. Normally distributed random numbers for Eq. \eqref{eqn:nondimS} are generated by applying a Box-Muller transform to uniformally distributed random numbers from the 64-bit linear congruential generator implemented in SPRNG-1.0 \cite{mascagni2000algorithm}. $\text{E}[{R_i}(t)]$ is trivially calculated as the mean grain size at the current timestep. We solve Eq. \eqref{eqn:nondimRm} implicitly for $R_m(t)$ by Newton's method, using the value of $R_m(t)$ at the previous timestep as an initial guess. Finally, we calculate $\Omega(t)$ by requiring the small-sample statistics of $\Omega(t)$, given in Eq. \eqref{eqn:st_statistics}, to hold at the end of an Euler-Maruyama timestep (see Appendix 1). Consequently, $\Omega(t)$ is given by
\begin{align}
    \Omega(t)^2 = \frac{1}{\Delta t} \left[ \frac{1}{n} \text{Var}[{R_i}(t)] - \left(1 - \frac{\Delta t}{ \text{E}[{R_i}]^2} \right)^2 \text{Var}[{S_i}(t)] \right] . \label{eqn:OMEGA_CALCULATION}
\end{align}

Grain growth necessarily involves the elimination of many grains as their volume is redistributed to the remaining grains. Simulation of an $({R_i}, {S_i})$ pair ceases if ${R_i}$ becomes less than or equal to zero. To maintain sufficient active grain realizations to construct a reliable histogram at the end of the simulation, the size distribution of active grains is re-sampled with replacement if the fraction of active grains drops below some specified fraction of the initial number of realizations. Since each realization is sensitive only to itself and the statistics of the global population of grains, re-sampling does not affect the overall results.

The starting grain sizes are drawn from a normal distribution with unit mean. Across a suite of simulations, we alter the initial variance of grain size between $\sim10^{-4}$ and $\sim10^{-1}$ to investigate the evolution toward the self-similar state. We use a time step $\Delta t = 10^{-5}$, which was determined by testing for converged statistics of the global set of grains (see Appendix 2).  Assuming that a grain may only interact with its neighbours, the number of grains within each local environment $n$ should be approximately the average number of faces of a grain plus one, to account for the grain itself. For space filling tetrakaidecahedra $n$ is $15$ \cite{thomson1887lxiii} and for random Voronoi polyhedra $n$ is $16.5$ \cite{wakai2000three}. We take $n = 16$ as a reference case. 

\section{Results}

The only adjustable parameter in our model is the number of grains within each local environment $n$. First, we explore the kinetics and grain-size distributions that result if $n$ is set to the reference value of $16$. Then we explore the effect of $n$ on the predicted grain-size distributions, using a suite of simulations with values of $n$ between $4$ and $64$, which all begin with the same normal distribution of grain size with unit mean and variance $0.03$.

\begin{figure}
  \centering
    \includegraphics[width=0.6\textwidth]{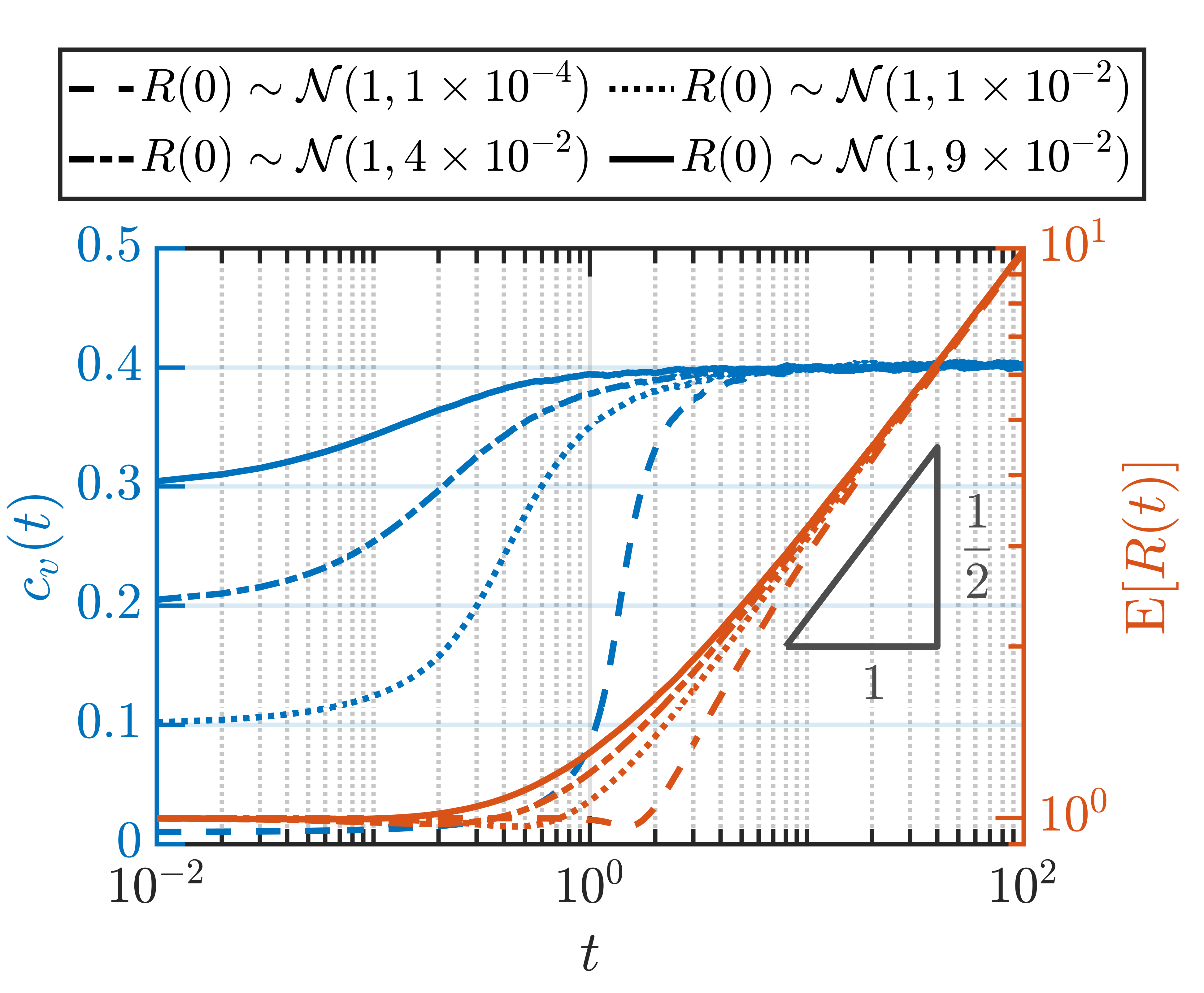}
    \caption{Time evolution of the coefficient of variation (left axis, blue) and mean grain size (right axis, red) for different starting distributions. These simulations were performed with $n=16$. The coefficient of variation converges to a value of 0.4, indicating that self-similarity is achieved. Once self-similarity of the grain-size statistics are achieved, the mean grain size converges to a square root of time relationship.}
    \label{fig:time_series}
\end{figure}

A key test of our model is whether it evolves to a self-similar state. If the simulation achieves self-similarity, then normalised statistics such as the coefficient of variation should be constant and histograms of grain size normalised by the mean should be independent of time. The coefficient of variation, plotted in Figure \ref{fig:time_series}, converges to $c_v = 0.40$ by $t=5$ for all initial conditions used, indicating that self-similarity is achieved and is stable---at least over the range of model times considered. This value is slightly higher than that obtained by Hillert ($c_v = 0.35$) \cite{hillert1965}. The time taken in the transient towards self-similarity changes with the initial variance of the grain-size distribution. Tighter initial distributions require a longer transient to reach self-similarity. Other normalised statistics also converge to a constant value. For example, the ratio of the mean-field radius to the mean grain size converges to $R_m/\text{E}[{R_i}] = 1.2$, which is higher than the value of 1.125 obtained in a pure Hillert model.

\begin{figure}
  \centering
    \includegraphics[width=0.6\textwidth]{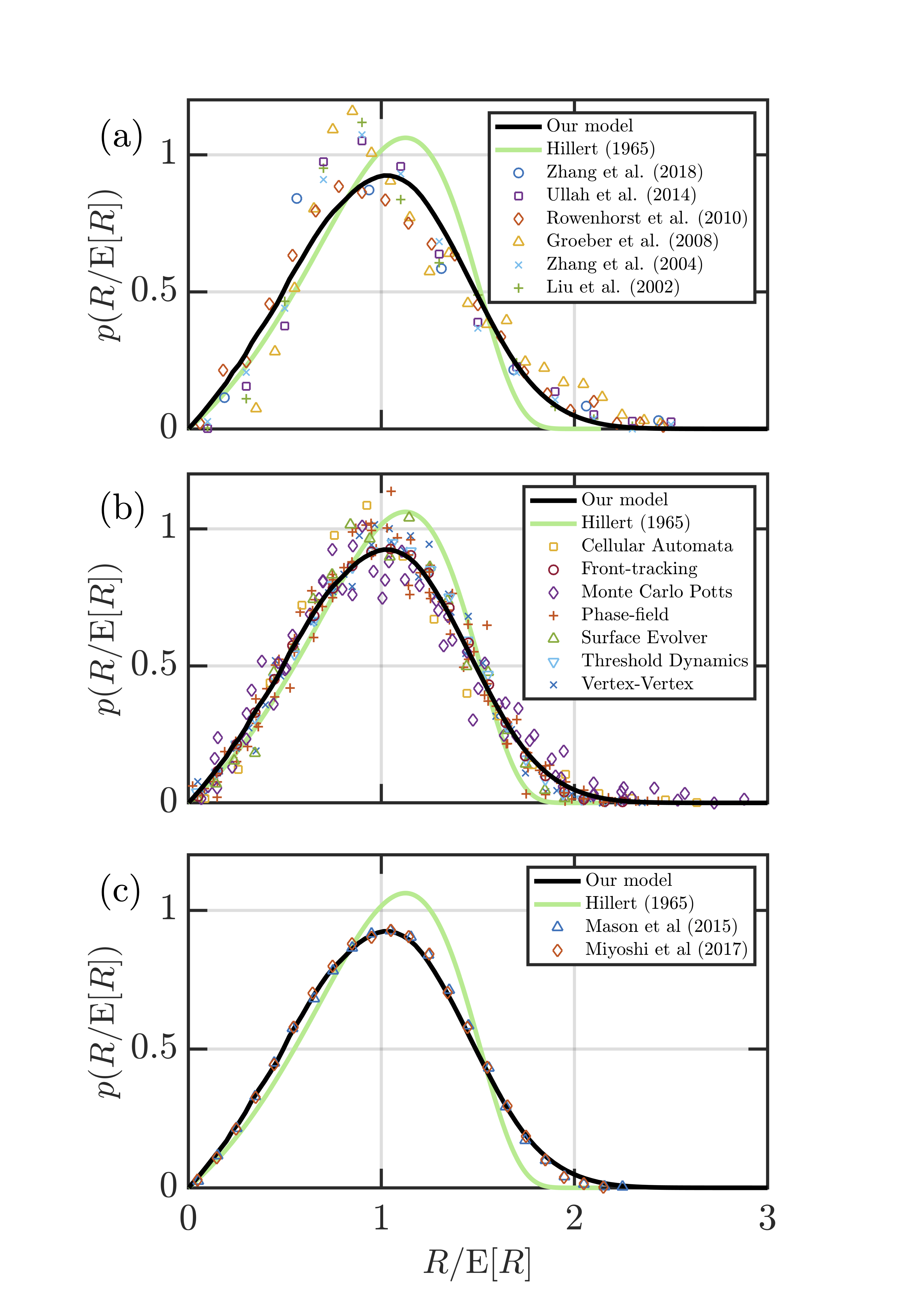}
    \caption{Comparison of the predicted probability distribution of grain size normalised by the mean in our model compared to the predictions of Hillert's model and observations from (a) physical experiments and (b and c) numerical simulations. Our distribution was obtained from a model with $n=16$. Experimental data is taken from \cite{liu2002three, zhang2004characterization, groeber2008framework, rowenhorst2010three, ullah2014three, zhang2018three}. The histograms of distributions from numerical simulations derive from a wide variety of approaches: the vertex-vertex method \cite{fuchizaki1995computer, weygand1999three}, the threshold dynamics method \cite{elsey2010large}, surface evolver \cite{wakai2000three}, the phase-field method \cite{krill2002computer, kamachali20123, kim2014phase, miyoshi2017ultra}, the Monte Carlo Potts method \cite{kim2005three, zollner2006three, ullah2017simulations}, cellular automata \cite{ding2006cellular} and a front tracking method \cite{mason2015geometric}. In (c) our predicted distribution is compared with the distributions from the state-of-the-art simulations of Mason et al \cite{mason2015geometric}, who use a front tracking algorithm, and Miyoshi et al \cite{miyoshi2017ultra}, who use the phase-field method. }
    \label{fig:data_comparison}
\end{figure}

Histograms of grain size normalised by the mean also converge to a self-similar state. This convergence occurs following an initial transient in all of our simulations during which the normalised distribution broadens. Figure \ref{fig:data_comparison} compares the distribution obtained under our model (with $n = 16$) with the histogram from a pure Hillert model. In contrast to the Hillert distribution that is left skewed, our distribution is approximately symmetrical about the mean. In addition, our distribution is broader than the Hillert distribution. In the Hillert distribution, 99.9\% of grains are smaller than $1.8 \text{E}[R_i]$, whereas in our distribution, the equivalent percentile is located at $2.2 \text{E}[R_i]$.

These differences from the prediction of the Hillert model mean that our distribution gives a better fit to  histograms of distributions from experiments and numerical simulations that resolve and evolve the grain structure. In Figure \ref{fig:data_comparison}, our distribution is compared to experimental data from iron \cite{zhang2004characterization, ullah2017simulations, zhang2018three}, steel \cite{liu2002three}, IN100 \cite{groeber2008framework}, and $\beta$-titanium \cite{rowenhorst2010three}, gathered by serial sectioning and diffraction contrast tomography, and data from numerical simulations applying the vertex-vertex method \cite{fuchizaki1995computer, weygand1999three}, the threshold dynamics method \cite{elsey2010large}, surface evolver \cite{wakai2000three}, the phase-field method \cite{krill2002computer, kamachali20123, kim2014phase, miyoshi2017ultra}, the Monte Carlo Potts method \cite{kim2005three, zollner2006three, ullah2017simulations}, cellular automata \cite{ding2006cellular} and a front tracking method \cite{mason2015geometric}. The mode of the experimental histograms is closer to the mode of our distribution than the Hillert distribution, and the tail of the experimental histograms compares favourably to our distribution. Our distribution lies within the envelope of histograms from numerical simulations; the mode and the tail of our distribution matches the histograms from numerical simulations. 

Finally, in Figure \ref{fig:data_comparison}c, we highlight the comparison with two state-of-the-art numerical simulations, performed by Mason et al \cite{mason2015geometric} and Miyoshi et al \cite{miyoshi2017ultra}. Mason et al. applied a front-tracking method that solves the MacPherson-Srolovitz equation \cite{macpherson2007neumann}, which expresses the rate of volume change in terms of topological information, to high accuracy.  Miyoshi et al. apply the phase-field method to a microstructure with a large number of grains and take care to ensure that a true steady-state distribution is obtained. These simulations provide the most accurate available estimates of the steady-state grain-size distribution developed in normal grain growth and hence are the best distributions against which to compare our results. Our distribution is in excellent agreement with the distributions from Mason et al. \cite{mason2015geometric} and Miyoshi et al \cite{miyoshi2017ultra}.

A fundamental observation from experiments and numerical simulations is that the mean grain size grows like the square root of time \cite{rios2018}. Under our model, the mean grain size, plotted in Figure \ref{fig:time_series} as a function of time, grows like the square root of time following an initial transient. The transient extends to $t \approx 10$, which is comparable to the time taken for the coefficient of variation to converge to its value in the self-similar state. The length of the transient in the evolution of mean grain size depends on the initial variance of the grain-size distribution; the lower the initial variance, the longer the transient. Following the transient, the mean grain size in all simulations is well approximated by
\begin{align}
    \text{E}[{R_i}(t)]^2 = a + b t.
\end{align}
The slope of this fit, $b = 0.98$, is greater than the value of $0.79$ obtained under a pure Hillert model, indicating that the growth rate of mean grain size is increased by accounting for the  heterogeneity in grain environments.

\begin{figure}
  \centering
    \includegraphics[width=0.6\textwidth]{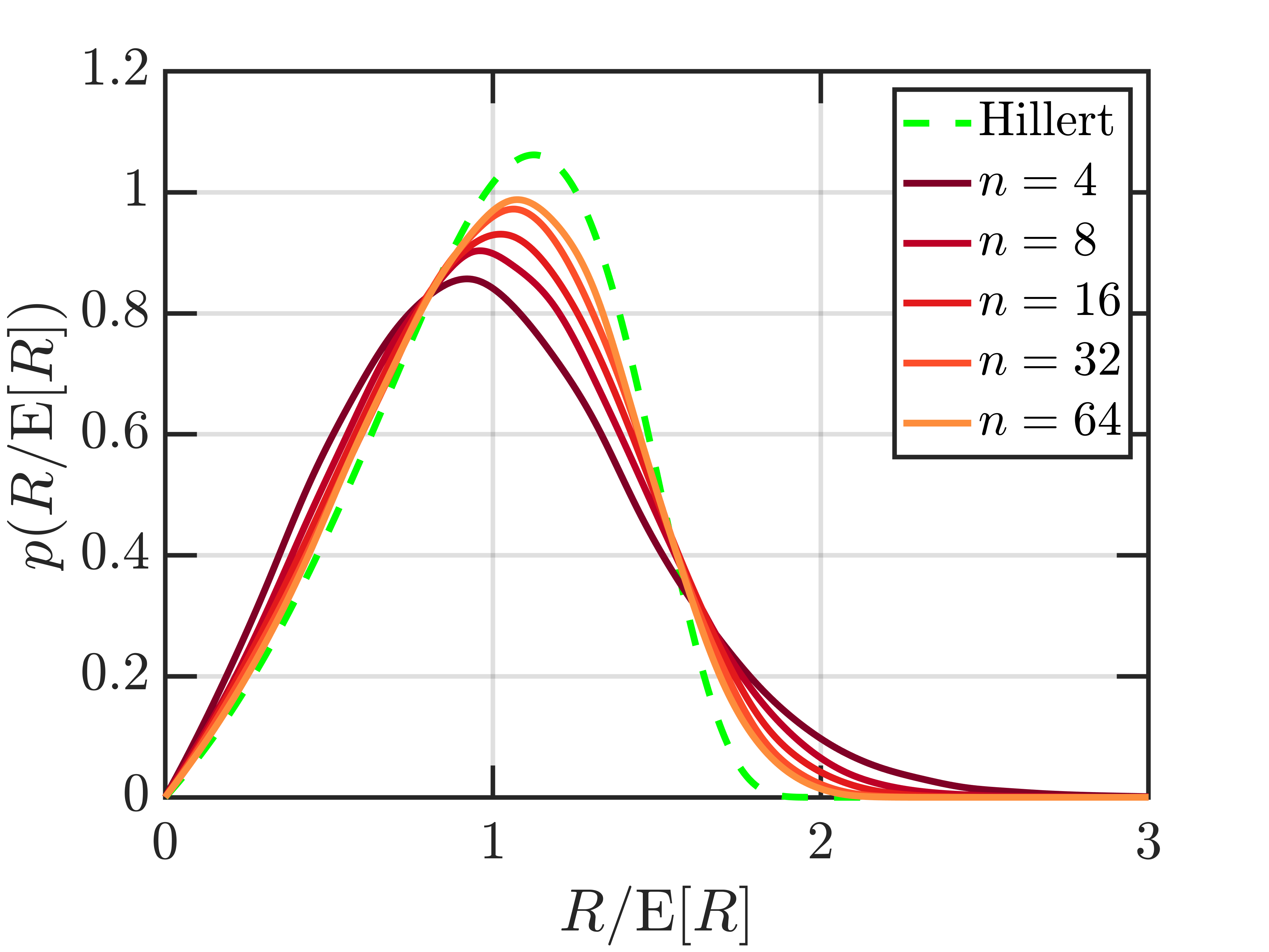}
    \caption{Normalised grain-size distributions from  simulations with different choices of the number of grains within a local environment $n$, compared to the Hillert distribution. The simulations are initialised with normally distributed grain size with unit mean and variance of 0.3, and run until the distribution of grain size reaches steady state. As $n$ increases, the distribution approaches the Hillert model.}
    \label{fig:variableN}
\end{figure}

Figure \ref{fig:variableN} compares normalised, self-similar grain-size distributions obtained from simulations with different choices of $n$. As $n$ increases, the mode of the distribution shifts to the right and the tail of the distribution shrinks, such that the distribution moves closer to the Hillert distribution. In the limit of large $n$, where the local environment encompasses the total population of grains, our model reduces to the Hillert model and so the Hillert distribution is expected. Hence the self-similar distributions that result from our model form a family that depend on the number of grains within the local environment and approach the Hillert distribution as $n$ increases.

\section{Discussion}

\subsection{Self-similar distribution}

Our model, in which the Hillert theory is augmented with a random variable describing each grain's local environment, reproduces the key observations of normal grain growth: square-root of time kinetics and a self-similar distribution of grain size. The grain-size distribution it produces, with  $n=16$ chosen by analogy with space-filling polyhedra, is in better agreement with distributions from experimental data than the Hillert distribution and is in good agreement with distributions from numerical simulations. Furthermore, our distribution is in excellent agreement with the distributions derived by the state-of-the-art numerical simulations of Mason et al. \cite{mason2015geometric} and Miyoshi et al. \cite{miyoshi2017ultra}. The improved agreement with experimentally derived histograms suggests that accounting for heterogeneity in grain environments represents the leading-order correction to Hillert's model.
 
Notwithstanding this improvement over the Hillert distribution, the mode of the experimental distributions is smaller than the mode obtained under our model. This inconsistency could result from systematic bias present in experimental distributions: smaller grains may be under-counted because of finite image resolution or due to the serial sectioning process applied to obtain three-dimensional information. As the distributions are normalised by an overestimated mean, this bias will act to change the shape of normalised histograms of grain size at small grain sizes and shift the mode of these histograms to smaller values. Furthermore, whilst experimental investigations attempt to isolate normal grain growth, there may be a number of processes acting in real materials that modify the kinetics of grain-size evolution away from idealised models. For example, the effects of anisotropy in grain-surface energy or mobility \cite[e.g.,][]{kazaryan2002grain}, solute drag \cite[e.g.,][]{kim2008grain}, pore drag \cite[e.g.,][]{karato1989grain}, and Zener pinning by unintended secondary phases \cite[e.g.,][]{nes1985zener} may act to modify grain-growth kinetics and consequently modify the distribution of grain size. As numerical simulations exclude these processes, they present a more direct test of our hypothesis.

\begin{figure}
  \centering
    \includegraphics[width=0.6\textwidth]{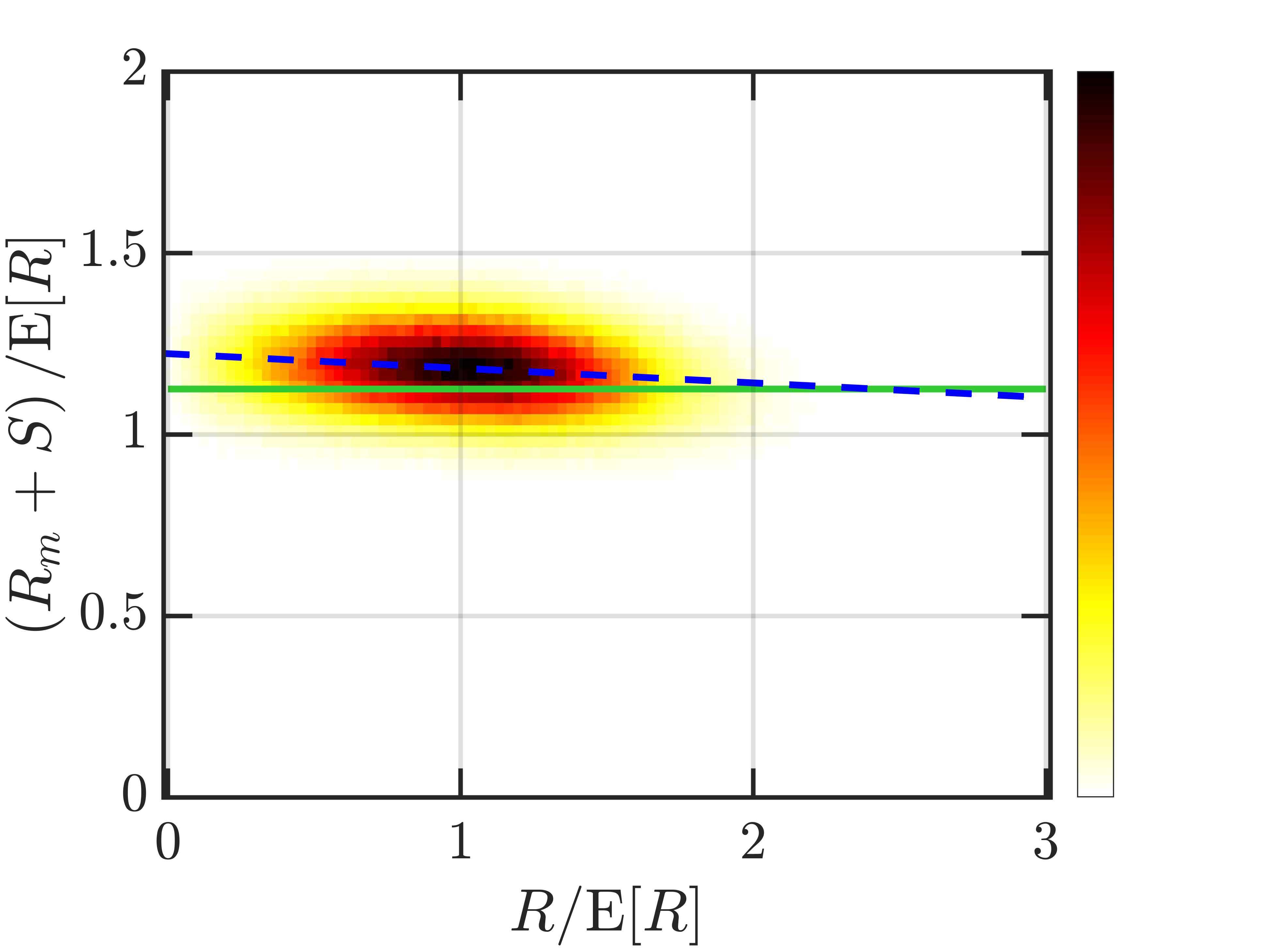}
    \caption{Heat map of grain size vs. local environment radius for the results of a simulation that achieved a self-similar state. In this simulation, $n$ was equal to 16. The mean-field radius under the Hillert model is plotted as a solid green line. A linear fit is plotted as a dashed blue line. The heat map colour indicates the relative density of points, with white indicating low density and black indicating high density.}
    \label{fig:r_vs_s}
\end{figure}

A potential explanation for the broader shape of our distribution is that a correlation develops between $R_i$ and $S_i$ such that our model is effectively a correlation-field theory, similar to that proposed by Marthinsen et al. \cite{marthinsen1996influence}. A correlation may develop under our model because grains with relatively large local environments are more likely to shrink in size. The relationship between local environment radius and grain size is explored in Figure \ref{fig:r_vs_s}. A weak relationship exists between grain size and local environment radius, with the same sense as the correlation that Marthinsen et al. \cite{marthinsen1996influence} observed in their Monte Carlo Potts model. However, local environment radius varies little with grain size compared to the variance of local environment radius at any given grain size. Our results thus indicate that the correlation between grain size and local environment is not a significant control on normal grain growth.

\begin{figure}
  \centering
    \includegraphics[width=0.6\textwidth]{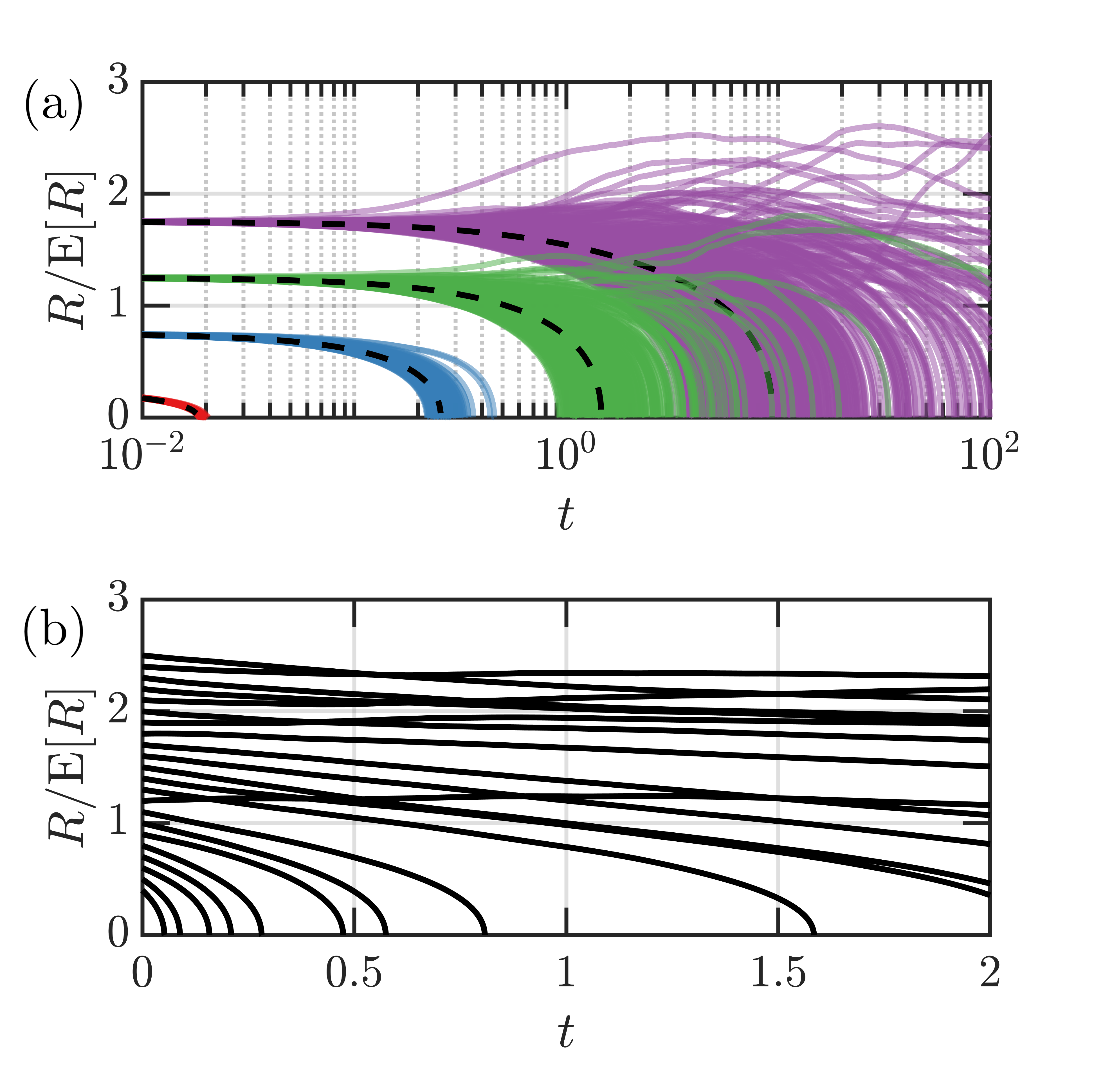}
    \caption{
    (a) Realizations of $R_i(t)$ normalised by the mean grain size compared to trajectories obtained under the Hillert model. For the simulation of these trajectories, $n$ was taken to be equal to 16. Time $t$ is relative to an initially self-similar state such that the trajectories are representative of self-similar evolution. Coloured lines show the evolution under our model for four different initial grain sizes $R_i(0) = 0.25, 0.75, 1.25, 1.75$. Hillert trajectories are plotted as black-dashed lines. For grain sizes that are small relative to the mean, the local environment effect is small and trajectories follow the trajectory of the Hillert model. For larger grains, the local environment effect may dominate and trajectories deviate from the Hillert trend. A logarithmic scale is used to highlight behaviour of trajectories to high model times. (b) Set of realisations of ${R_i}$ normalised by the mean grain size once a self-similar distribution has been obtained, for $n= 16$. Compare to Figure 7b of Srolovitz et al. \cite{srolovitz1984computer}.}
    \label{fig:trajectories}
\end{figure}

The true role of the local environment variable can be investigated by examining the grain-size trajectories of individual grains, plotted in Figure \ref{fig:trajectories}. For grains that are initially smaller than the mean, the local environment radius has little effect; instead the self-curvature term in Eq. \eqref{eqn:nondimR} dominates the evolution of grain size. For grains that are larger than the mean, the local environment radius dominates the evolution of grain size and a grain's self-curvature is unimportant. Since the local environment radius performs a random walk, it causes grains that are larger than the mean to also undergo a random walk. This mixing of deterministic evolution, for grains smaller than the mean, and random perturbations, for grains larger than the mean, matches the observations made by Srolovitz \cite{srolovitz1984computer} of grain-size trajectories in a Monte Carlo simulation of grain growth (compare Figure 7b of \cite{srolovitz1984computer} with our Figure \ref{fig:trajectories}b).

In the Hillert model, the stability argument applied in the derivation of the grain-size distribution imposes an upper cut-off on its extent. This cut-off limits the range of the Hillert distribution and leads to the discrepancy between the Hillert distribution and observations. In our model, larger grains perform a random walk, and so the distribution extends to higher values of normalised grain size, leading to better agreement with distributions from experiments and numerical simulations.

\subsection{Effect of local environment size}

Simulations that explore the effect of the local environment size indicate that as $n$ increases, the self-similar distribution approaches the Hillert distribution (Figure \ref{fig:variableN}). Within the Hillert model, all grains in the global set form part of the local environment of any given grain. In our model, this corresponds to the limit of large $n$, where $n$ becomes equal to the total number of grains considered. Mathematically, as $n$ becomes large, the variance of $S_i$ given by Eq. \eqref{eqn:st_statistics} becomes small and so the magnitude of $S_i$ shrinks towards zero. In the case that $S_i$ is negligible, our Eq. \eqref{eqn:our_model} reduces to the Hillert model. Hillert's model therefore represents an asymptotic limit of our model for large $n$.

The family of distributions generated by varying $n$ can be compared against previously published distributions. The distributions derived by Rios \cite{rios1999comparisonBROWN} agree well with results from experiments and computer simulations \cite{rios2006comparison}. Rios obtained a  one-parameter family of distributions that depend on a parameter $\nu$, with $\nu = 4$ corresponding to the Hillert distribution. Indeed, the Rios distribution arises by relaxing a stability argument that Hillert applied to constrain $\nu$ to be 4. Values of $\nu$ between 2 and 3.6 produce distributions that compare favourably against observed distributions \cite{rios2006self}.

To compare against the Rios distributions, we determine the value of $\nu$ that generates the Rios distribution that most closely approximates our own distribution for a given $n$.  To measure the closeness of two distributions, we define the statistical divergence between our model and the Rios distribution in terms of the total variation distance as
\begin{align}
D(\nu, n) = \frac{1}{2} \int \left| p(x; n) - q(x; \nu) \right| \text{d}x,
\label{eqn:stdisim}
\end{align}

where $D(\nu, n)$ is the statistical divergence, $x$ is the grain size normalised by the mean, $p(x; n)$ is our own distribution for a given $n$ and $q(x; \nu)$ is the Rios distribution as defined in \cite{rios2006comparison} for a given $\nu$. The total variation distance is zero for distributions that match perfectly and is unity for distributions that do not overlap. We approximate the integral in Eq. \eqref{eqn:stdisim} using the trapezium rule and minimise $D(\nu, n)$ using MATLAB's {\it fminsearch} function to find the value $\nu = \nu^*$ that best approximates our own distribution at a fixed value of $n$. 

\begin{figure}[htbp]
  \centering
    \includegraphics[width=0.6\textwidth]{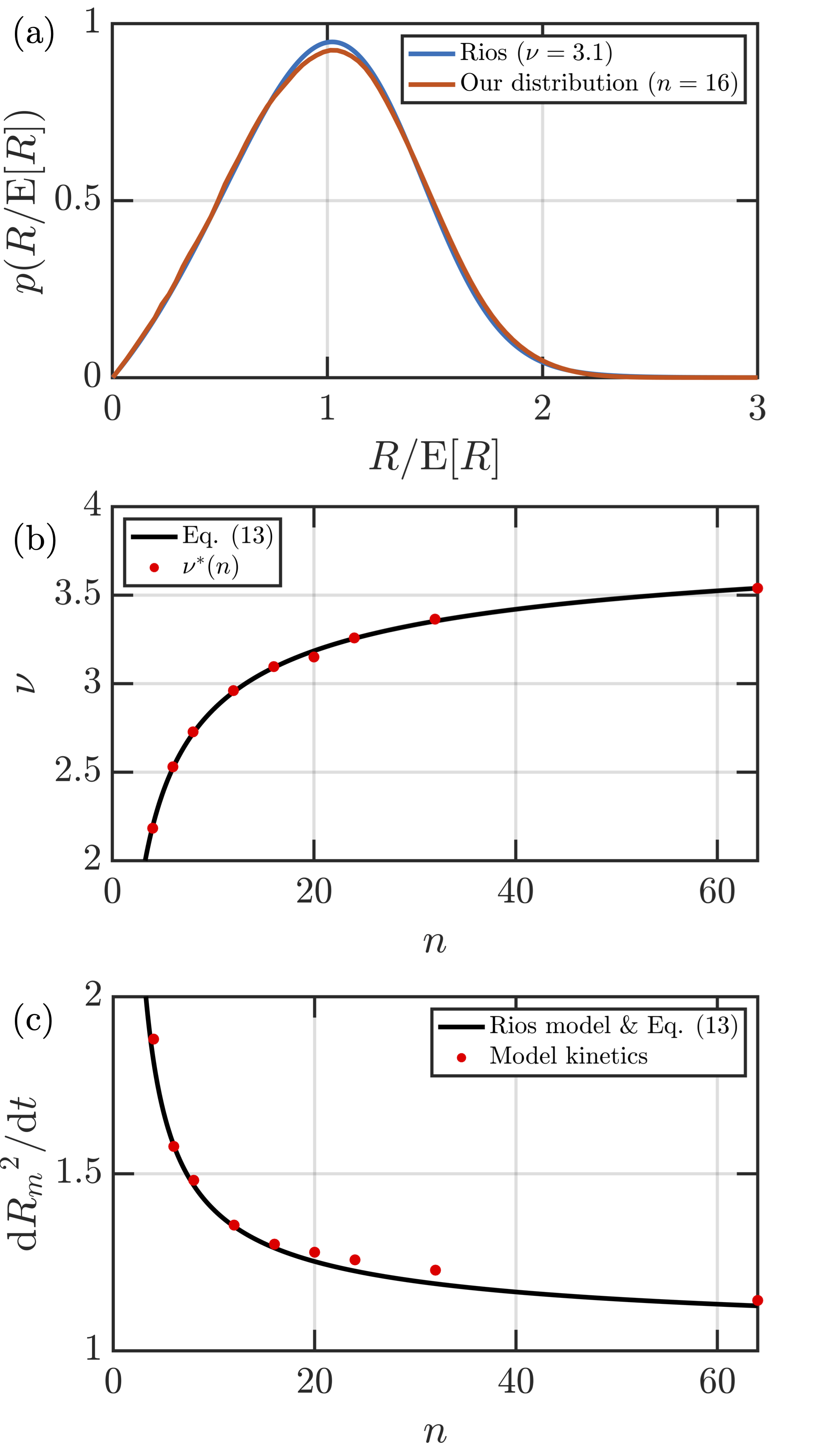}
    \caption{(a) Comparison of the distribution obtained under our model with $n=16$ and the most closely matching Rios distribution with $\nu = 3.1$. (b) The parameter $\nu*$ that generates the best-fitting Rios distribution against the size of the local environment $n$.
    The black line is a fit to $\nu*$ given by Eq. \eqref{eqn:approxnun}. (c) The rate of change of the square of the mean-field-radius against local environment $n$. The predicted kinetics from Rios et al. \cite{rios2006comparison} are recast in terms of the local environment $n$ using Eq. \eqref{eqn:approxnun} (black line). The kinetics observed (red dots), calculated using a finite-difference approximation across a time step, match the predicted kinetics. }
    \label{fig:n16Rios}
\end{figure}

For all of the values of $n$ that we explore, there is a Rios distribution that matches our distribution. In each case, the minimised total variation distance is below 3\%, indicating a correspondence between the two distributions. Figure 6a presents a comparison between our distribution for $n=16$ and the best-fit Rios distribution, which has $\nu^*$ equal to 3.1. The relationship between the number of grains in a local environment $n$ and the value of $\nu^*$ that generates the best-fitting Rios distribution is explored in Figure 6b. This relationship is approximated by

\begin{align}
\nu^* = 4 - 3.6/\sqrt{n}.
\label{eqn:approxnun}
\end{align}
This correspondence can be further explored by examining the steady-state grain-size kinetics. In Rios et al.'s \cite{rios2006comparison} approach, the rate of change of the square of the mean-field radius depends inversely on the parameter $\nu$. We recast this prediction to depend on the local environment $n$ using Eq. \eqref{eqn:approxnun} and compare it against the steady-state grain-size kinetics observed in our simulations in Figure \ref{fig:n16Rios}c. The observed kinetics closely match the predicted kinetics derived using Eq. \eqref{eqn:approxnun}, further highlighting the  correspondence between the steady-state results of our stochastic model and the predictions of the phenomenological model of Rios et al. \cite{rios2006comparison}. This correspondence between the distributions and the kinetics of the mean-field-radius suggests that $\nu$ might be interpreted as representing the average number of grains adjacent to each individual grain. This interpretation is a testable prediction of our model.

\subsection{Model limitations}

A critical assumption applied in our model is that the sampling distribution used to describe the probability density of the local environment radius $R_m + S_i$ can be approximated with a normal distribution. We test this assumption by subsampling the grains from one simulation to construct grain subsets, taking $n$ to be equal to 16. We then calculate the local environment radius of each subset as ${\text{E}_i}[{R_i}^2]/{\text{E}_i}[{R}_i]$. This quantity is indeed normally distributed, with the statistics given by our approximation in Eq. \eqref{eqn:st_statistics} (see Appendix 3). In general, the accuracy of this approximation will depend on the underlying distribution of grain size that is being sampled and the size of the local environment $n$, which determines the number of samples in each subset. Although our normal approximation for the sampling distribution holds well for the grain-size distributions we have considered, it may break down for arbitrary grain-size distributions far from steady state. In such cases, our model would not be applicable.

Another critical assumption is that the number of grains comprising any particular local environment is constant and equal to $n$. We have selected a reasonable value of $n$ based on space-filling polyhedra. However, in practice, the number of faces and therefore the number of neighbours of a grain depends on the size of the grain, due to a size--topology correlation. Smaller grains should, in general, have fewer neighbours than larger grains and therefore smaller neighbourhoods. Consequently, the variance in the local environment radius ought to be higher for smaller grains. Based on the observed size--topology relationships \cite[e.g.,][]{kamachali20123}, we expect nearly an order of magnitude decrease in the variance of the local environment radius from the smallest to the largest grains. However, the impact of heterogeneity is not felt equally by all grains. As is evident in Figure \ref{fig:trajectories}, the evolution of smaller grains is dominated by the self-curvature term in Eq. \eqref{eqn:our_model}; the effect of the local environment only becomes important once the grain size exceeds the mean. The neighbourhoods of the largest grains are expected to be approximately twice as big as the neighbourhoods of grains whose size is equal to the mean \cite{kamachali20123}, leading to only a factor of two difference in the variance of the local environment radius between these two cases. In principle, the correlation between neighbourhood size $n$ and grain size $R_i$ should influence the statistics of the local environment radius. However, over the range of grains for which the local environment radius matters, the impact of the size--topology correlation is limited. Future work could incorporate the dependence of neighbourhood size on grain size by altering the type of noise introduced by $S$ to depend on the size of each grain directly.

\subsection{Source of fluctuations}

The source of fluctuations in stochastic models is typically assumed to be topological in nature \cite{pande2020beyond, mullins1998grain}. If the full topological state of a three-dimensional grain is known, then the rate of change of its volume can be calculated exactly using the MacPherson-Srolovitz relationship \cite{macpherson2007neumann}, which is deterministic. MacPherson \& Srolovitz derive an approximation to this relationship, in which the rate of change of size of a grain is proportional to the difference between the square root of the number of its faces and a constant \cite{macpherson2007neumann} (see also a similar relationship due to Rios \& Glicksman \cite{rios2006self}). Events that modify the topological properties of grains are expected to drive discrete changes in the growth rate of grains through this relationship. Such events are therefore a candidate source of noise in normal grain growth. For example, grain disappearances can alter the number of faces possessed by surrounding grains. Within our model, noise is calibrated to represent heterogeneity in the local environment of grains, which may be alternatively understood in terms of grain topology. The number of faces of a grain depends, in an average sense, on its relative size: smaller grains are expected to have fewer faces. Similarly, grains that are surrounded by smaller-than-average grains would be expected to have a greater number of faces. In our model, the local environment radius provides an average measure of the size of grains surrounding a given grain and therefore holds information that could be used to infer the topological properties of the grain. Consequently, local environments can be understood in topological terms; heterogeneity in local environments can be identified with heterogeneity in topological properties. This reasoning can be applied to the trajectories obtained under our model. In Figure \ref{fig:trajectories}b, there exists a grain trajectory that does not follow the (Hillert-like) trend of the grains that have a similar size. Instead of shrinking relative to the mean, its size relative to the mean grain size remains approximately constant. The reason for this is that it has a smaller local environment radius than the average, which stabilises it according to Eq. \eqref{eqn:our_model}. From the topological perspective, we can infer that due to its comparatively smaller local environment radius, the neighbouring grains are smaller, and so the number of faces that our grain possesses is correspondingly larger than would otherwise be expected for a grain of its size. According to the approximate MacPherson-Srolovitz equation, its volume (and therefore radius) is stabilised by its topology, consistent with our results.

The noise supplied to our model by the Ornstein-Uhlenbeck process for $S_i$ differs from the white noise incorporated into previous stochastic models of normal grain growth \cite{pande1990n2, pande2008grain, pande2010self, pande2020beyond}. A key distinction lies in the time-correlation of the noise. As the noise applied in previous models is white noise, the time-correlation function in these models is a Dirac delta. In contrast, in our model noise is supplied to the grain-size kinetics by the local environment radius. The local environment radius is generated by an Ornstein-Uhlenbeck process that is correlated on a timescale $\tau$ as a result of the drift term in Eq. \eqref{eqn:SDE_St}. We have chosen this timescale to be the macroscopic timescale of normal grain growth. The fundamental assumption involved in the application of white noise is that the `collisions' (in analogy to Brownian motion) driving the stochastic component are both fast and uncorrelated. These collisions are understood to act on a microscopic timescale that is much smaller than the macroscopic timescale, such that in the limit of the microscopic timescale tending to zero, a Dirac delta time correlation function is obtained. Mullins \cite{mullins1998grain} argued that the events constituting the collisions of normal grain growth would occur on the macroscopic timescale, and so could not give rise to a fluctuation term described by white noise. This critique does not apply to our model in which the noise supplied is correlated on the macroscopic timescale, and so can reasonably represent the random influence of the events described by Mullins. 

Pande \& Moser \cite{pande2020beyond} argue against Mullins's criticism of the application of a white noise term by applying a more global perspective. They assert that each collision event that drives the stochastic component of normal grain growth affects the entire set of grains. Consequently, as the number of grains in the system becomes large, the timescale for an individual event that may affect all grains becomes small, justifying the introduction of a white noise term. This justification cannot apply to the stochastic component of our model, as we have emphasised a more local perspective. In our model, grains are screened from the global environment and are influenced only by the grains with which they can exchange mass. Consequently, the influence of any given event is more limited. This difference in perspective between local and global scales constitutes a key distinction between our model and previous stochastic models.

There are other distinctions between our approach and previous stochastic models that apply white noise to the grain-growth rate. The amplitude of noise in our model is constrained by association with a specific hypothesis about heterogeneity in local environments. In contrast, in models that apply white noise, the amplitude of the white noise is a free parameter because no individual physical process is directly identified as the key contributor to the noise term. Given these differences between our model and other stochastic models of normal grain growth, it is surprising that the results of both approaches seem to be equivalent. Indeed, Pande \& Moser \cite{pande2020beyond} compare the approximate solutions found by Pande \& McFadden \cite{pande2010self} for their white-noise-driven model to the distributions obtained by the phase-field simulation of Miyoshi et al.'s \cite{miyoshi2017ultra}, the experimentally-derived distribution of Rowenhorst et al. \cite{rowenhorst2010three} and the phenomenological distribution due to Rios \cite{rios2006comparison}, as we have done. Pande \& Moser find that their distribution is in agreement with the distributions obtained by Miyoshi et al. and Rowenhorst et al., and is congruent with the Rios distribution, as we have found with our distribution. Furthermore, as Pande \& Moser \cite{pande2020beyond} note, the approximate grain-size distribution derived from their white-noise-driven model is analytically equivalent to the grain-size distribution that Streitenberger \& Z\"{o}llner \cite{streitenberger2006effective} derived from their modified mean-field theory, which they calibrated against observations from simulations \cite{zollner2006three}. The close correspondence between the distributions obtained here, white-noise-driven models, and modified mean-field theories raises the question. Why do these different approaches result in such similar grain-size distributions?

\section{Conclusions}

By accounting for heterogeneity in the local environment of grains and its resultant effect on mass exchange between grains, our model resolves the discrepancy between the Hillert distribution and observations. It also matches the key characteristics of normal grain growth. The improved fit is achieved because heterogeneity in local environments causes relatively large grains to undergo a random walk in grain size, allowing some of these grains to grow larger than is possible under the Hillert model. Our distribution is therefore broader than Hillert's and provides a better match to observed distributions. It is in excellent agreement with distributions from state-of-the-art simulations. Our analysis suggests a physical meaning for the fitting parameter in the Rios distribution in terms of the size of the local environment.

The noise we have introduced into our model is fundamentally different to the unconstrained white noise typically introduced into stochastic models of normal grain growth. The amplitude of the noise in our model is determined by association with a hypothesis of heterogeneity in the local environments of grains. This hypothesis can be alternatively understood in terms of heterogeneity in grain topology. Furthermore, the noise in our model is correlated on the timescale of normal grain growth. It is therefore interesting that the grain-size distribution obtained by our model is similar to the distributions that develop in these other stochastic models.

\section{Acknowledgements}

We would like to thank the reviewers for their comments, which have improved the manuscript. The authors acknowledge the use of the University of Oxford Advanced Research Computing (ARC) facility in carrying out this work (http://dx.doi.org/10.5281/zenodo.22558). This research received funding from the European Research Council under Horizon 2020 research and innovation program grant agreement number 772255.
T.B. was supported by a Natural Environment Research Council (NERC) studentship in the Oxford NERC Doctoral Training Partnership. S.T. acknowledges a Research Fellowship from All Souls College, Oxford.

\section{Appendix 1: Calculation of $\Omega(t)$}

Requiring that the small sample statistics of ${S}_i$ hold at the end of a timestep implies that 
\begin{align}
    \text{Var}[{S_i}(t + \Delta t)] = \frac{1}{n} \text{Var}[{R_i}(t)].
    \label{eqn:omegad1} \tag{A.1}
\end{align}
Under the Euler-Maruyama scheme, Eq. \eqref{eqn:nondimS} gives
\begin{align}
    {S_i}(t + \Delta t) = \left( 1 - \frac{\Delta t}{\text{E}[{R}_i(t)]^2} \right) {S}_i(t) + \Omega(t) \sqrt{\Delta t} \mathcal{N}(0, 1). \tag{A.2}
\end{align}
Consequently, the variance of ${S}_i$ following a timestep is given by
\begin{align}
    \text{Var}[{S}_i(t + \Delta t)] = \left(1 - \frac{\Delta t}{\text{E}[{R_i}]^2} \right)^2 \text{Var}[{S}_i(t)] + \Omega(t)^2 \Delta t .
    \label{eqn:omegad2} \tag{A.3}
\end{align}
By equating \eqref{eqn:omegad1} and \eqref{eqn:omegad2}, we can determine $\Omega(t)$ at any timestep such that Eq. \eqref{eqn:st_statistics} holds,
\begin{align}
    \Omega(t)^2 = \left[ \frac{1}{n} \text{Var}[{R}_i(t)] - \left(1 - \frac{\Delta t}{\text{E}[{R}_i]^2} \right)^2 \text{Var}[{S_i}(t)] \right] / \Delta t, \tag{A.4}
\end{align}
which is Eq. \eqref{eqn:OMEGA_CALCULATION} in the main text.

\section{Appendix 2: Choice of $\Delta t$}

\begin{figure}[htbp]
  \centering
    \includegraphics[width=0.6\textwidth]{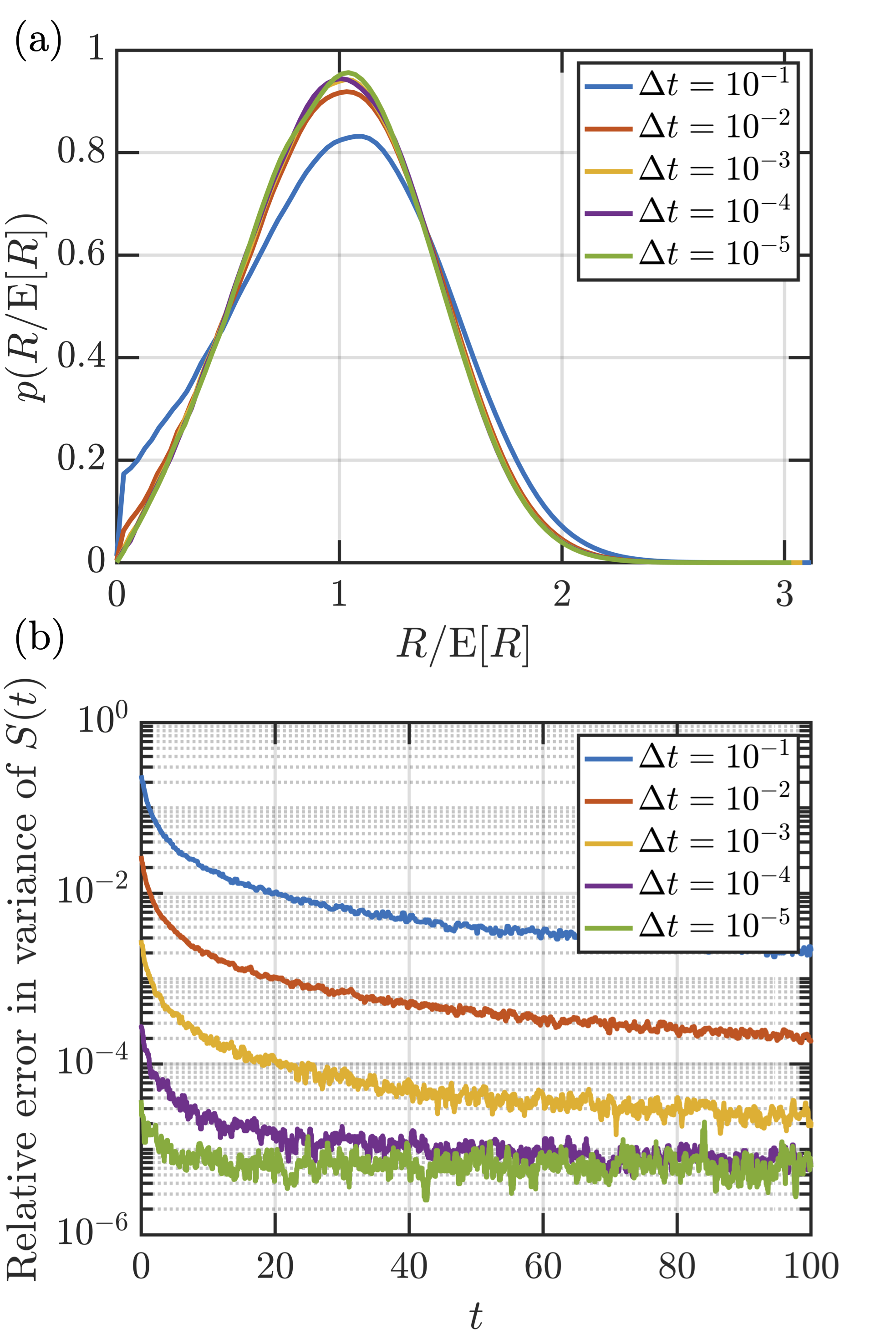}
    \caption{(a) Histograms of grain size normalised by the mean, obtained from models run with different choices of time step $\Delta t$ (see legend). All model runs were started with the same initial distribution of grain size (normally distributed with unit mean and variance of 0.03).    (b) Time series of relative error in the variance of $S(t)$ from models with different choices of time step $\Delta t$. All model runs were started with a normal distribution of grain size with unit mean and variance of 0.03. The relative error in variance of $S(t)$ was calculated as the difference between the expected variance of $S(t)$ from Eq. \eqref{eqn:st_statistics} and observed variance of $S(t)$ normalised by the expected variance.}
    \label{fig:dtchoice}
\end{figure}

One method of choosing the appropriate time step $\Delta t$ is to compare histograms of grain size normalised by the mean produced by the model with different choices of $\Delta t$. Figure \ref{fig:dtchoice}a shows histograms of grain size normalised by the mean at $t = 1$, with the same initial condition of normally distributed grain size with mean equal to one and initial variance of 0.3, for a range of possible time steps. The histogram with $\Delta t = 0.1$ differs significantly from the other histograms, with a relative excess of grains of size close to zero. For $\Delta t = 0.01$ and smaller, the histogram  converges to a shape independent of the choice of $\Delta t$, indicating that a $\Delta t$ smaller than $10^{-3}$ is be appropriate.

Another method of evaluating the choice of $\Delta t$ is to assess the relative error between the assumed statistics of $S_i$ given by Eq. \eqref{eqn:st_statistics} and the observed statistics of $S$ within the model. Time series of the relative error in the variance of $S$ for different choices of $\Delta t$ are plotted in Figure \ref{fig:dtchoice}b. The magnitude of the relative error decreases with decreasing $\Delta t$ until $\Delta t = 10^{-5}$, after which decreasing the time step does not significantly reduce the magnitude of relative error, indicating that an appropriate choice of time step is $10^{-5}$.

\pagebreak
\section{Appendix 3: Assessing the assumption that $S$ is normally distributed}

\begin{figure}[htbp]
  \centering
    \includegraphics[width=0.6\textwidth]{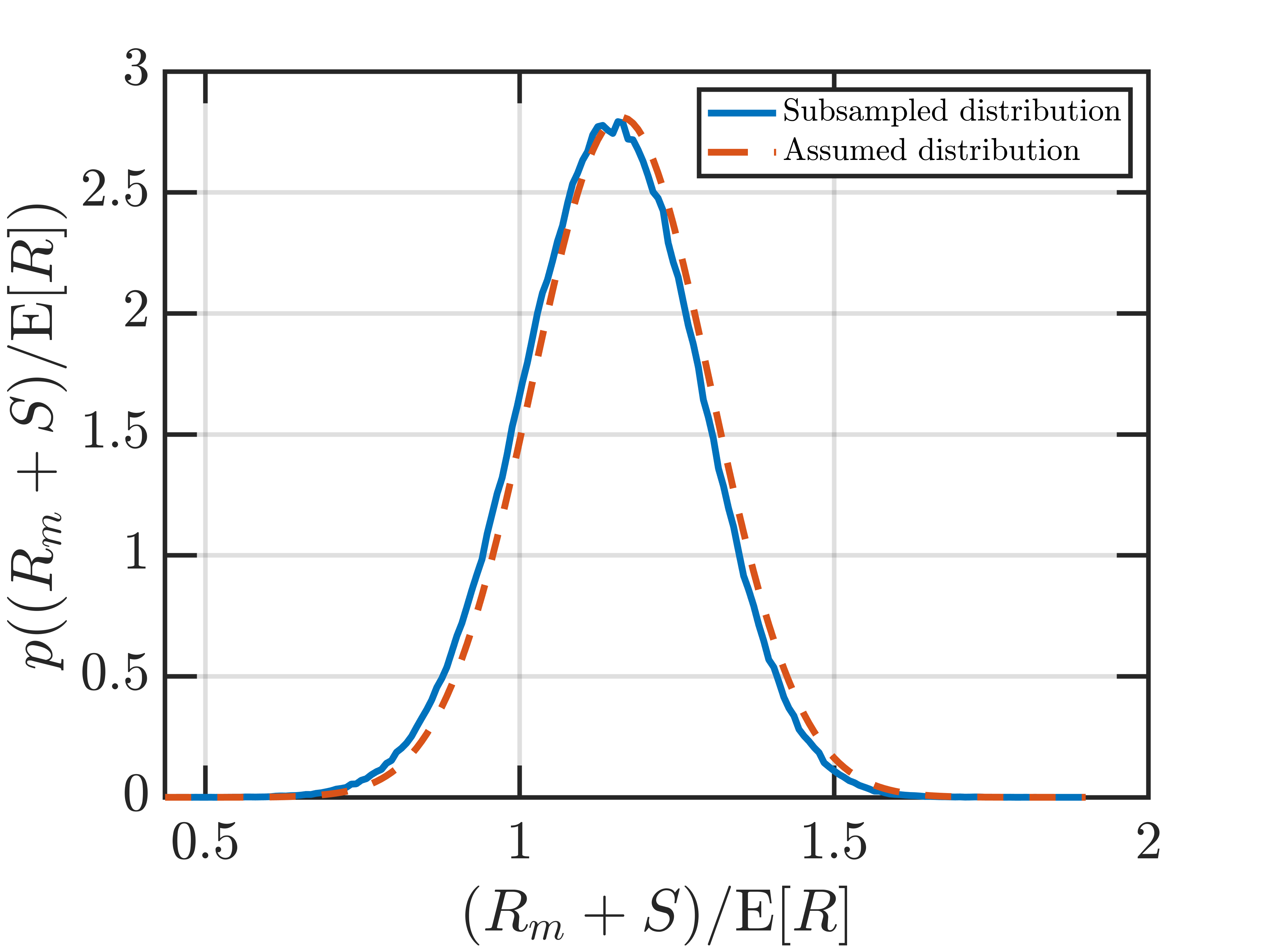}
    \caption{Comparison of the assumed distribution of $S$ given by Eq. \eqref{eqn:st_statistics} and the distribution obtained by subsampling a converged model to produce $10^6$ sets of local environments containing $n = $ 16 samples, and calculating the local environment radius directly.}
    \label{fig:subsampledS}
\end{figure}

The small-sample statistics of the local environment radius can be assessed by subsampling with replacement the observed grain sizes $R_i$ from the results of a model. The $R_i$ resulting from a single model run that converged to steady state are subsampled with replacement to produce $10^6$ sets of $16$ samples. The local environment radius $R_m + S_i$ is then calculated for each set of $n = 16$ samples according to ${\text{E}_i}[{R_i}^2]/{\text{E}_i}[{R_i}]$. The histogram of the local environment radius so calculated is compared in Figure \ref{fig:subsampledS} to the distribution of $R_m + S_i$, where $S_i$ is assumed to follow the statistics given by Eq. \eqref{eqn:st_statistics}. The distribution obtained by subsampling $R_i$ agrees well with the assumed distribution, and the relative error in the mean and variance between the distributions is less than a percent. This validates the approximation used in Eq. \eqref{eqn:st_statistics}.

\pagebreak
\bibliography{grain_pdf.bib}

\begin{thebibliography}{53}
\providecommand{\natexlab}[1]{#1}
\providecommand{\url}[1]{\texttt{#1}}
\expandafter\ifx\csname urlstyle\endcsname\relax
  \providecommand{\doi}[1]{doi: #1}\else
  \providecommand{\doi}{doi: \begingroup \urlstyle{rm}\Url}\fi

\bibitem[Rozel et~al.(2011)Rozel, Ricard, and
  Bercovici]{rozel2011thermodynamically}
Antoine Rozel, Yanick Ricard, and David Bercovici.
\newblock A thermodynamically self-consistent damage equation for grain size
  evolution during dynamic recrystallization.
\newblock \emph{Geophysical Journal International}, 184\penalty0 (2):\penalty0
  719--728, 2011.

\bibitem[Rios and Z{\"o}llner(2018)]{rios2018}
PR~Rios and D~Z{\"o}llner.
\newblock Critical assessment 30: Grain growth--unresolved issues.
\newblock \emph{Materials Science and Technology}, 34\penalty0 (6):\penalty0
  629--638, 2018.

\bibitem[Hillert(1965)]{hillert1965}
M~Hillert.
\newblock On the theory of normal and abnormal grain growth.
\newblock \emph{Acta metallurgica}, 13\penalty0 (3):\penalty0 227--238, 1965.

\bibitem[Rios et~al.(2006)Rios, Dalpian, Brandao, Castro, and
  Oliveira]{rios2006comparison}
PR~Rios, TG~Dalpian, VS~Brandao, JA~Castro, and ACL Oliveira.
\newblock Comparison of analytical grain size distributions with
  three-dimensional computer simulations and experimental data.
\newblock \emph{Scripta materialia}, 54\penalty0 (9):\penalty0 1633--1637,
  2006.

\bibitem[Rios and Glicksman(2006{\natexlab{a}})]{rios2006self}
Paulo~R Rios and Martin~E Glicksman.
\newblock Self-similar evolution of network structures.
\newblock \emph{Acta materialia}, 54\penalty0 (4):\penalty0 1041--1051,
  2006{\natexlab{a}}.

\bibitem[Fischer et~al.(2003)Fischer, Svoboda, and
  Fratzl]{fischer2003thermodynamic}
FD~Fischer, J~Svoboda, and P~Fratzl.
\newblock A thermodynamic approach to grain growth and coarsening.
\newblock \emph{Philosophical Magazine}, 83\penalty0 (9):\penalty0 1075--1093,
  2003.

\bibitem[Kertsch and Helm(2016)]{kertsch2016modelling}
Lukas Kertsch and Dirk Helm.
\newblock Modelling grain growth in the framework of rational extended
  thermodynamics.
\newblock \emph{Modelling and Simulation in Materials Science and Engineering},
  24\penalty0 (4):\penalty0 045001, 2016.

\bibitem[Rios and Glicksman(2006{\natexlab{b}})]{rios2006topological}
Paulo~R Rios and Martin~E Glicksman.
\newblock Topological theory of abnormal grain growth.
\newblock \emph{Acta materialia}, 54\penalty0 (19):\penalty0 5313--5321,
  2006{\natexlab{b}}.

\bibitem[Brown(1989)]{brown1989new}
LC~Brown.
\newblock A new examination of classical coarsening theory.
\newblock \emph{Acta metallurgica}, 37\penalty0 (1):\penalty0 71--77, 1989.

\bibitem[Pande and Dantsker(1990)]{pande1990n2}
CS~Pande and E~Dantsker.
\newblock On a stochastic theory of grain growth---{II}.
\newblock \emph{Acta Metallurgica et Materialia}, 38\penalty0 (6):\penalty0
  945--951, 1990.

\bibitem[Marthinsen et~al.(1996)Marthinsen, Hunderi, and
  Ryum]{marthinsen1996influence}
K~Marthinsen, O~Hunderi, and N~Ryum.
\newblock The influence of spatial grain size correlation and topology on
  normal grain growth in two dimensions.
\newblock \emph{Acta materialia}, 44\penalty0 (4):\penalty0 1681--1689, 1996.

\bibitem[Z{\"o}llner and Streitenberger(2006)]{zollner2006three}
Dana Z{\"o}llner and Peter Streitenberger.
\newblock Three-dimensional normal grain growth: {Monte Carlo Potts} model
  simulation and analytical mean field theory.
\newblock \emph{Scripta materialia}, 54\penalty0 (9):\penalty0 1697--1702,
  2006.

\bibitem[Pande and McFadden(2010)]{pande2010self}
CS~Pande and GB~McFadden.
\newblock Self-similar grain size distribution in three dimensions: A
  stochastic treatment.
\newblock \emph{Acta Materialia}, 58\penalty0 (3):\penalty0 1037--1044, 2010.

\bibitem[Pande and Moser(2020)]{pande2020beyond}
CS~Pande and AE~Moser.
\newblock Beyond modified mean field: a case for a stochastic grain growth
  model.
\newblock \emph{Philosophical Magazine}, 100\penalty0 (7):\penalty0 837--856,
  2020.

\bibitem[Uhlenbeck and Ornstein(1930)]{uhlenbeck1930theory}
George~E Uhlenbeck and Leonard~S Ornstein.
\newblock On the theory of the brownian motion.
\newblock \emph{Physical review}, 36\penalty0 (5):\penalty0 823, 1930.

\bibitem[Wang and Uhlenbeck(1945)]{wang1945theory}
Ming~Chen Wang and George~Eugene Uhlenbeck.
\newblock On the theory of the brownian motion ii.
\newblock \emph{Reviews of modern physics}, 17\penalty0 (2-3):\penalty0 323,
  1945.

\bibitem[Rios(1999)]{rios1999comparisonBROWN}
PR~Rios.
\newblock Comparison between a computer simulated and an analytical grain size
  distribution.
\newblock \emph{Scripta materialia}, 40\penalty0 (6):\penalty0 665--668, 1999.

\bibitem[Rios and L{\"u}cke(2001)]{rios2001comparison}
P.R Rios and K~L{\"u}cke.
\newblock Comparison of statistical analytical theories of grain growth.
\newblock \emph{Scripta Materialia}, 44\penalty0 (10):\penalty0 2471 -- 2475,
  2001.

\bibitem[Streitenberger(1998)]{streitenberger1998generalized}
P~Streitenberger.
\newblock Generalized lifshitz-slyozov theory of grain and particle coarsening
  for arbitrary cut-off parameter.
\newblock \emph{Scripta materialia}, 39\penalty0 (12):\penalty0 1719--1724,
  1998.

\bibitem[Pande and Rajagopal(2001)]{pande2001uniqueness}
CS~Pande and AK~Rajagopal.
\newblock Uniqueness and self similarity of size distributions in grain growth
  and coarsening.
\newblock \emph{Acta materialia}, 49\penalty0 (10):\penalty0 1805--1811, 2001.

\bibitem[Pande et~al.(2008)Pande, Cooper, and McFadden]{pande2008grain}
CS~Pande, KP~Cooper, and GB~McFadden.
\newblock Grain size distribution in two dimensions in the long time limit.
\newblock \emph{Acta materialia}, 56\penalty0 (18):\penalty0 5304--5311, 2008.

\bibitem[Pande and Dantsker(1991)]{pande1991n3}
CS~Pande and E~Dantsker.
\newblock On a stochastic theory of grain growth---{III}.
\newblock \emph{Acta metallurgica et materialia}, 39\penalty0 (6):\penalty0
  1359--1365, 1991.

\bibitem[Streitenberger and Z{\"o}llner(2006)]{streitenberger2006effective}
P~Streitenberger and D~Z{\"o}llner.
\newblock Effective growth law from three-dimensional grain growth simulations
  and new analytical grain size distribution.
\newblock \emph{Scripta materialia}, 55\penalty0 (5):\penalty0 461--464, 2006.

\bibitem[Mullins(1998)]{mullins1998grain}
WW~Mullins.
\newblock Grain growth of uniform boundaries with scaling.
\newblock \emph{Acta materialia}, 46\penalty0 (17):\penalty0 6219--6226, 1998.

\bibitem[Rios(2004)]{rios2004irreversible}
PR~Rios.
\newblock Irreversible thermodynamics, parabolic law and self-similar state in
  grain growth.
\newblock \emph{Acta materialia}, 52\penalty0 (1):\penalty0 249--256, 2004.

\bibitem[Balay et~al.(2018)Balay, Abhyankar, Adams, Brown, Brune, Buschelman,
  Dalcin, Eijkhout, Gropp, Kaushik, Knepley, May, McInnes, Mills, Munson, Rupp,
  Sanan, Smith, Zampini, Zhang, and Zhang]{petsc-user-ref}
Satish Balay, Shrirang Abhyankar, Mark~F. Adams, Jed Brown, Peter Brune, Kris
  Buschelman, Lisandro Dalcin, Victor Eijkhout, William~D. Gropp, Dinesh
  Kaushik, Matthew~G. Knepley, Dave~A. May, Lois~Curfman McInnes, Richard~Tran
  Mills, Todd Munson, Karl Rupp, Patrick Sanan, Barry~F. Smith, Stefano
  Zampini, Hong Zhang, and Hong Zhang.
\newblock {PETS}c users manual.
\newblock Technical Report ANL-95/11 - Revision 3.9, Argonne National
  Laboratory, 2018.

\bibitem[Balay et~al.(1997)Balay, Gropp, McInnes, and Smith]{petsc-efficient}
Satish Balay, William~D. Gropp, Lois~Curfman McInnes, and Barry~F. Smith.
\newblock Efficient management of parallelism in object oriented numerical
  software libraries.
\newblock In E.~Arge, A.~M. Bruaset, and H.~P. Langtangen, editors,
  \emph{Modern Software Tools in Scientific Computing}, pages 163--202.
  Birkh{\"{a}}user Press, 1997.

\bibitem[Mascagni and Srinivasan(2000)]{mascagni2000algorithm}
Michael Mascagni and Ashok Srinivasan.
\newblock Algorithm 806: {SPRNG}: A scalable library for pseudorandom number
  generation.
\newblock \emph{ACM Transactions on Mathematical Software (TOMS)}, 26\penalty0
  (3):\penalty0 436--461, 2000.

\bibitem[Thomson(1887)]{thomson1887lxiii}
William Thomson.
\newblock {LXIII. On} the division of space with minimum partitional area.
\newblock \emph{The London, Edinburgh, and Dublin Philosophical Magazine and
  Journal of Science}, 24\penalty0 (151):\penalty0 503--514, 1887.

\bibitem[Wakai et~al.(2000)Wakai, Enomoto, and Ogawa]{wakai2000three}
Fumihiro Wakai, Naoya Enomoto, and Hiroshi Ogawa.
\newblock Three-dimensional microstructural evolution in ideal grain
  growth---general statistics.
\newblock \emph{Acta Materialia}, 48\penalty0 (6):\penalty0 1297--1311, 2000.

\bibitem[Liu et~al.(2002)Liu, Yu, and Qin]{liu2002three}
Guoquan Liu, Haibo Yu, and Xiangge Qin.
\newblock Three-dimensional grain topology--size relationships in a real
  metallic polycrystal compared with theoretical models.
\newblock \emph{Materials Science and Engineering: A}, 326\penalty0
  (2):\penalty0 276--281, 2002.

\bibitem[Zhang et~al.(2004)Zhang, Enomoto, Suzuki, and
  Ishimaru]{zhang2004characterization}
C~Zhang, M~Enomoto, A~Suzuki, and T~Ishimaru.
\newblock Characterization of three-dimensional grain structure in
  polycrystalline iron by serial sectioning.
\newblock \emph{Metallurgical and Materials Transactions A}, 35\penalty0
  (7):\penalty0 1927--1933, 2004.

\bibitem[Groeber et~al.(2008)Groeber, Ghosh, Uchic, and
  Dimiduk]{groeber2008framework}
Michael Groeber, Somnath Ghosh, Michael~D Uchic, and Dennis~M Dimiduk.
\newblock A framework for automated analysis and simulation of 3d
  polycrystalline microstructures.: Part 1: Statistical characterization.
\newblock \emph{Acta Materialia}, 56\penalty0 (6):\penalty0 1257--1273, 2008.

\bibitem[Rowenhorst et~al.(2010)Rowenhorst, Lewis, and
  Spanos]{rowenhorst2010three}
DJ~Rowenhorst, AC~Lewis, and G~Spanos.
\newblock Three-dimensional analysis of grain topology and interface curvature
  in a $\beta$-titanium alloy.
\newblock \emph{Acta Materialia}, 58\penalty0 (16):\penalty0 5511--5519, 2010.

\bibitem[Ullah et~al.(2014)Ullah, Liu, Luan, Li, ur~Rahman, and
  Ali]{ullah2014three}
Asad Ullah, Guoquan Liu, Junhua Luan, Wenwen Li, Mujeeb ur~Rahman, and Murad
  Ali.
\newblock Three-dimensional visualization and quantitative characterization of
  grains in polycrystalline iron.
\newblock \emph{Materials Characterization}, 91:\penalty0 65--75, 2014.

\bibitem[Zhang et~al.(2018)Zhang, Zhang, Ludwig, Rowenhorst, Voorhees, and
  Poulsen]{zhang2018three}
Jin Zhang, Yubin Zhang, Wolfgang Ludwig, David Rowenhorst, Peter~W Voorhees,
  and Henning~F Poulsen.
\newblock Three-dimensional grain growth in pure iron. {Part I}. {S}tatistics
  on the grain level.
\newblock \emph{Acta Materialia}, 156:\penalty0 76--85, 2018.

\bibitem[Fuchizaki et~al.(1995)Fuchizaki, Kusaba, and
  Kawasaki]{fuchizaki1995computer}
Kazuhiro Fuchizaki, Takuo Kusaba, and Kyozi Kawasaki.
\newblock Computer modelling of three-dimensional cellular pattern growth.
\newblock \emph{Philosophical Magazine B}, 71\penalty0 (3):\penalty0 333--357,
  1995.

\bibitem[Weygand et~al.(1999)Weygand, Br{\'e}chet, L{\'e}pinoux, and
  Gust]{weygand1999three}
D~Weygand, Y~Br{\'e}chet, J~L{\'e}pinoux, and W~Gust.
\newblock Three-dimensional grain growth: a vertex dynamics simulation.
\newblock \emph{Philosophical Magazine B}, 79\penalty0 (5):\penalty0 703--716,
  1999.

\bibitem[Elsey et~al.(2010)Elsey, Esedoglu, and Smereka]{elsey2010large}
Matt Elsey, Selim Esedoglu, and Peter Smereka.
\newblock Large-scale simulation of normal grain growth via diffusion-generated
  motion.
\newblock \emph{Proceedings of the Royal Society A: Mathematical, Physical and
  Engineering Sciences}, 467\penalty0 (2126):\penalty0 381--401, 2010.

\bibitem[Krill~Iii and Chen(2002)]{krill2002computer}
CE~Krill~Iii and L-Q Chen.
\newblock Computer simulation of 3-d grain growth using a phase-field model.
\newblock \emph{Acta materialia}, 50\penalty0 (12):\penalty0 3059--3075, 2002.

\bibitem[Kamachali and Steinbach(2012)]{kamachali20123}
Reza~Darvishi Kamachali and Ingo Steinbach.
\newblock 3-d phase-field simulation of grain growth: Topological analysis
  versus mean-field approximations.
\newblock \emph{Acta Materialia}, 60\penalty0 (6-7):\penalty0 2719--2728, 2012.

\bibitem[Kim et~al.(2014)Kim, Kim, Dong, Steinbach, and Lee]{kim2014phase}
Hyun-Kyu Kim, Seong~Gyoon Kim, Weiping Dong, Ingo Steinbach, and Byeong-Joo
  Lee.
\newblock Phase-field modeling for 3d grain growth based on a grain boundary
  energy database.
\newblock \emph{Modelling and Simulation in Materials Science and Engineering},
  22\penalty0 (3):\penalty0 034004, 2014.

\bibitem[Miyoshi et~al.(2017)Miyoshi, Takaki, Ohno, Shibuta, Sakane,
  Shimokawabe, and Aoki]{miyoshi2017ultra}
Eisuke Miyoshi, Tomohiro Takaki, Munekazu Ohno, Yasushi Shibuta, Shinji Sakane,
  Takashi Shimokawabe, and Takayuki Aoki.
\newblock Ultra-large-scale phase-field simulation study of ideal grain growth.
\newblock \emph{NPJ Computational Materials}, 3\penalty0 (1):\penalty0 1--6,
  2017.

\bibitem[Kim et~al.(2005)Kim, Hwang, Kim, Kwun, and Chae]{kim2005three}
YJ~Kim, SK~Hwang, MH~Kim, SI~Kwun, and Soo~Won Chae.
\newblock Three-dimensional {Monte-Carlo} simulation of grain growth using
  triangular lattice.
\newblock \emph{Materials Science and Engineering: A}, 408\penalty0
  (1-2):\penalty0 110--120, 2005.

\bibitem[Ullah et~al.(2017)Ullah, Khan, Weihua, Hussain, ur~Rahman, Salamat,
  Haq, et~al.]{ullah2017simulations}
Asad Ullah, Matiullah Khan, Xue Weihua, Safdar Hussain, Mujeeb ur~Rahman,
  Nadeem Salamat, Fazal Haq, et~al.
\newblock Simulations of grain growth in realistic {3D} polycrystalline
  microstructures and the {MacPherson--Srolovitz} equation.
\newblock \emph{Materials Research Express}, 4\penalty0 (6):\penalty0 066502,
  2017.

\bibitem[Ding et~al.(2006)Ding, He, Liu, and Ding]{ding2006cellular}
HL~Ding, YZ~He, LF~Liu, and WJ~Ding.
\newblock Cellular automata simulation of grain growth in three dimensions
  based on the lowest-energy principle.
\newblock \emph{Journal of Crystal Growth}, 293\penalty0 (2):\penalty0
  489--497, 2006.

\bibitem[Mason et~al.(2015)Mason, Lazar, MacPherson, and
  Srolovitz]{mason2015geometric}
Jeremy~K Mason, Emanuel~A Lazar, Robert~D MacPherson, and David~J Srolovitz.
\newblock Geometric and topological properties of the canonical grain-growth
  microstructure.
\newblock \emph{Physical Review E}, 92\penalty0 (6):\penalty0 063308, 2015.

\bibitem[MacPherson and Srolovitz(2007)]{macpherson2007neumann}
Robert~D MacPherson and David~J Srolovitz.
\newblock The von neumann relation generalized to coarsening of
  three-dimensional microstructures.
\newblock \emph{Nature}, 446\penalty0 (7139):\penalty0 1053--1055, 2007.

\bibitem[Kazaryan et~al.(2002)Kazaryan, Wang, Dregia, and
  Patton]{kazaryan2002grain}
A~Kazaryan, Y~Wang, SA~Dregia, and BR~Patton.
\newblock Grain growth in anisotropic systems: comparison of effects of energy
  and mobility.
\newblock \emph{Acta Materialia}, 50\penalty0 (10):\penalty0 2491--2502, 2002.

\bibitem[Kim and Park(2008)]{kim2008grain}
Seong~Gyoon Kim and Yong~Bum Park.
\newblock Grain boundary segregation, solute drag and abnormal grain growth.
\newblock \emph{Acta Materialia}, 56\penalty0 (15):\penalty0 3739--3753, 2008.

\bibitem[Karato(1989)]{karato1989grain}
SI~Karato.
\newblock Grain growth kinetics in olivine aggregates.
\newblock \emph{Tectonophysics}, 168\penalty0 (4):\penalty0 255--273, 1989.

\bibitem[Nes et~al.(1985)Nes, Ryum, and Hunderi]{nes1985zener}
E~Nes, N~Ryum, and O~Hunderi.
\newblock On the zener drag.
\newblock \emph{Acta Metallurgica}, 33\penalty0 (1):\penalty0 11--22, 1985.

\bibitem[Srolovitz et~al.(1984)Srolovitz, Anderson, Sahni, and
  Grest]{srolovitz1984computer}
DJ~Srolovitz, Michael~P Anderson, Paramdeep~S Sahni, and Gary~S Grest.
\newblock Computer simulation of grain growth---{II}. {Grain} size
  distribution, topology, and local dynamics.
\newblock \emph{Acta metallurgica}, 32\penalty0 (5):\penalty0 793--802, 1984.

\end{thebibliography}

\end{document}